# Aerospace Human System Integration Evolution over the Last 40 Years

Guy André Boy, Ph.D., FlexTech Chair Institute Professor
CentraleSupélec (Paris Saclay University) & ESTIA Institute of Technology
Chair, Human Systems Integration Working Group,
International Council on Systems Engineering
Chair, Aerospace Technical Committee, International Ergonomics Association
guy-andre.boy@centralesupelec.fr
+33 6 73 11 79 38

## Abstract

This chapter focuses on the evolution of Human-Centered Design (HCD) in aerospace systems over the last forty years. Human Factors and Ergonomics first shifted from the study of physical and medical issues to cognitive issues circa the 1980s. The advent of computers brought with it the development of human-computer interaction (HCI), which then expanded into the field of digital interaction design and User Experience (UX). We ended up with the concept of interactive cockpits, not because pilots interacted with mechanical things, but because they interacted using pointing devices on computer displays. Since the early 2000s, complexity and organizational issues gained prominence to the point that complex systems design and management found itself center stage, with the spotlight on the role of the human element and organizational setups. Today, Human Systems Integration (HSI) is no longer only a single-agent problem, but a multi-agent research field. Systems are systems of systems, considered as representations of people and machines. They are made of statically and dynamically articulated structures and functions. When they are at work, they are living organisms that generate emerging functions and structures that need to be considered in evolution (i.e., in their constant redesign). This chapter will more specifically, focus on human factors such as human-centered systemic representations, life critical systems, organizational issues, complexity management, modeling and simulation, flexibility, tangibility and autonomy. The discussion will be based on several examples in civil aviation and air combat, as well as aerospace.

## 1. Introduction

For the last forty years, aerospace systems evolved tremendously, mainly due to constant increasing automation, improvement of design and development methods and tools, and most importantly under the constant search for more safety, efficiency and usability. Air Traffic Management (ATM) is a matter of System of Systems (SoS), which requires a more solid systemic approach where technology, organizations and people are properly considered in an integrated framework. The search for such an approach is the main objective of this chapter. What are the purposeful attributes of ATM systems? What are purposeful nominal and off-nominal contexts of ATM systems? What are ATM disruptive factors? What are ATM intrinsic and extrinsic factors?

What is the best ATM topological model that can support analysis, design and evaluation of the growing real ATM system? What are the new ATM metrics in terms of operational performance, decision-making, trust and collaboration, for example? What is the role of Human-In-The-Loop Simulation (HITLS) in system-of-systems design?

This chapter first presents the evolution of aviation and the difficult issue of separability in the overall aeronautical community. Section 2 then discuss how software took the lead on hardware during the last fifty years or so, introducing a drastic shift from automation to autonomy in the aerospace domain. In addition to human models and related human-factors approaches, human roles drastically changed in the real world. Human models shifted from single-agent to multi-agent representations and related approaches. This evolution contributed to the emergence of new disciplines (section 3 of this chapter). Digitalization of industrial processes during the whole life cycle of products brought forward tangibility issues (section 4). The consideration of unexpected situations urges us to augment, and in some cases replace, rigid automation produced during the $20^{th}$ century by flexible autonomy (section 5). All these observations lead to the development of an appropriate human-centered systemic framework in the form of a conceptual framework, which will be useful for Human Systems Integration (HSI) (section 6). The conclusion will emphasize future endeavors of HSI as a generic process in our increasingly digitized society.

## 2. Evolution of aviation: A human systems integration perspective

At beginning of aviation, Air Traffic Controllers (ATCOs) had to guess both current and future aircraft positions in order to reduce uncertainty (Figure 1).

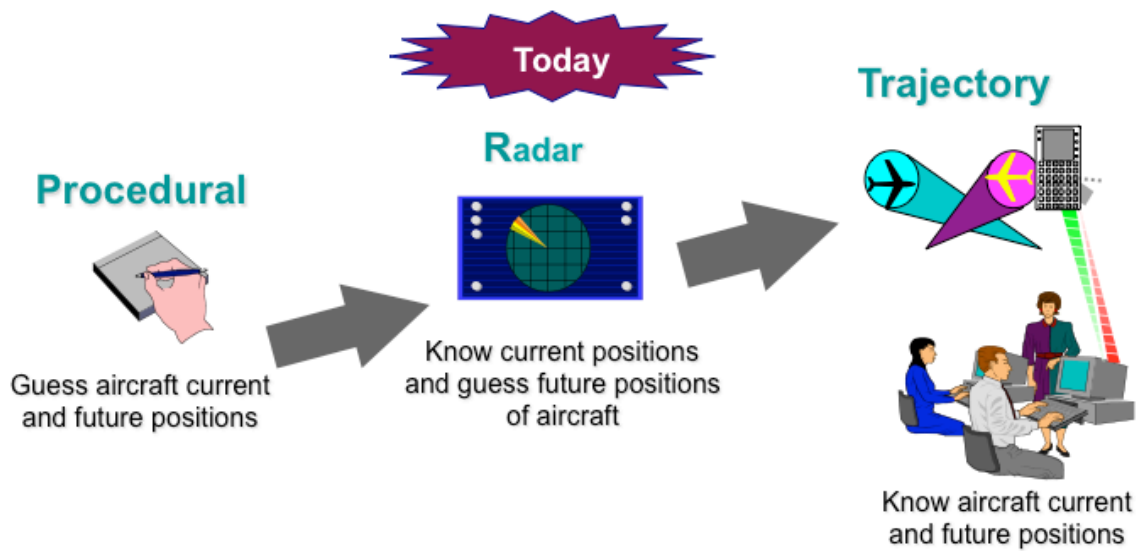

*Figure 1. ATC-to-ATM evolution from procedural control to trajectory management.*

During the second phase, based on the use of radar technology, ATCOs know aircraft positions but still have to guess future aircraft positions. We remain in this era despite saturation which in bustling airports demands shifting to the next phase, which is trajectory management, also called Trajectory-Based Operations (TBO). In TBO, ATCOs know both current and future aircraft positions, which is intended to reduce uncertainty considerably. While this looks great, a new problem emerges from the implementation of the TBO solution, which is necessary planning (e.g., 4D trajectories). Planning involves rigidity. Whenever rule-bound procedures are applicable, everything runs smoothly, but when unexpected situations occur, air traffic managers require flexibility. Rigidity and flexibility are contradictory concepts! This is the reason we now need to think in terms of flexible trajectory planning.

Another observation worth noting is that even if pilots and ATCOs are interacting at operations time, aircraft and ATC manufacturers and suppliers are very different institutions that rarely talk to each other. By contrast, Human-Centered Design (HCD) requires participatory design of complex systems (Boy, 2013). The space shuttle design and development is a good example of such an HCD approach where onboard and ground systems were designed and developed in concert. TBO requires such a multi-agent design approach. Methods were developed to this end (Boy, 1998, 2011), and the TOP model should be used to support the HCD of the overall aeronautical system (Figure 2) should be used toward the best articulation of Technology, Organizations and People.

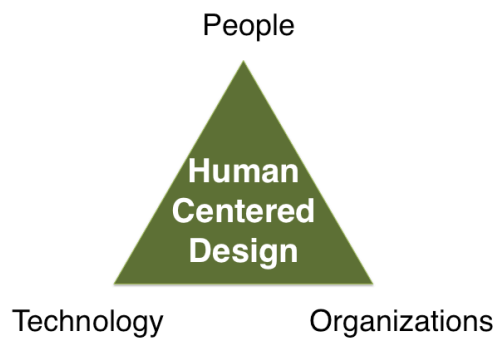

*Figure 2. The TOP Model (Boy, 2020).*

ATM complexity and automation are two major concepts in contemporary aviation. Aeronautical automation effectively started in commercial aviation in the 1930s (e.g., the Boeing 247 commercial aircraft flew with an autopilot in 1933). Aeronautics has then a long experience in automation. However, a big jump happened during the 1980s when glass cockpits started to be developed. This kind of digital automation drastically changed the way pilots interacted with their aircraft (Wiener, 1989). Technologically speaking, we moved from analog mechanical instruments to digital displays (e.g., primary flight display and navigation display) and controls (e.g., side sticks and auto-throttle). If the number of instruments in the cockpit drastically decreased, the quantity of information exponentially increased. A typical question is: how could such information be organized within the limited space provided by cockpit screens? Substantial research efforts have been carried out to answer this question (Boy, 1995; Doyon-Poulin et al., 2012; Letondal et al., 2018). More generally, complexity exponentially increased (i.e., the number of processed parameters and related systems increased), shifting from mechanical to digital complexity.

These digital systems were developed to take care of handling qualities, navigation and many other mechanical things. Systems invaded aircraft to the point that pilots ended up with a system of systems, not externally yet, but internally to the cockpit. Such a digitalization induced incremental production of layers and layers of electronics and software, added on top of each other, and resulting in the separation of pilots from the real physical world. For example, two decades ago, we started to talk about "interactive" cockpits, not because pilots interacted mechanically with engines, flaps and slats as they used to do since the beginning of aviation, but because they are now interacting with a pointing device on cockpit screens. We are talking about human-computer interaction in the cockpit (Boy, 1993).

In addition to cockpit automation (the technological side), the number of technical aircrew members in transoceanic flights was reduced over the last 60 years or so: five until the 1950s when the Radio Navigator was removed (the radio navigator was dedicated to voice communication equipment); four until the 1970s when the Navigator was removed (when inertial navigation systems were introduced); three until the 1980s when the Flight Engineer was removed (new monitoring equipment for engines and aircraft systems were introduced); and two to date. The next change will shortly happen if the Single Pilot Operations (SPO) goal is reached. Reducing crews involves organizational changes that need to be seriously considered, again in terms of safety, efficiency and comfort. At this point, it is crucial to understand whether such changes are evolutionary or revolutionary. Circa the end of the 1980s, we became conscious of this sociotechnical distinction when the Airbus 320 was certified. The A320 was highly automated compared to other commercial aircrafts of its category. Many experts thought that it was easier to fly. Some others claimed that such automation was dangerous. We did not realize at that time that even if we thought that automation was developed incrementally, almost linearly (we thought!), the nature of pilot's job radically changed from control to management (i.e., from control of aircraft trajectory to management of systems, which were controlling the trajectory). Once this job shift was understood, everything went right. This is the reason why systems engineering requires taking into account HSI since the beginning of the design and development process.

Historically, the design of a system was done in silos and, in many cases, systems were only connected just before operations, which is not a problem when sub-systems in the system are separable (Figure 3). Clumsy integration, often done too late in the development process, is likely to cause surprises and, sometimes, a few catastrophes. This is the reason why adjustments are always required, either operationally via adapted procedures and/or interfaces, and in the worst case more drastic redesign of the system itself. The separability concept has been used for a long time by physiologists to denote the possibility of separating an organ from the human body to work on it separately and put it back. Some organs (i.e., systems) are separable, that is the overall body (i.e., a system of systems) does not die from this momentary separation. Some other organs, such as the brain, cannot be separated because the human being could die from this separation. Therefore, those organs have to be investigated and treated while connected to the rest of the body.

In addition, 20$^{th}$ century engineering involved technicians who were, and still are, working in isolation and focused on a specific field or discipline. They were, and still are, barely aware of the integration of the overall complex SoS, i.e., "the whole picture" that they build. HSI has to consider interconnected SoSs, where systems include people and machines. This kind of requirement does not fit the current urgency of fast market economy… or fast anything for that matter! It takes time to get complex systems working well and maturing from three points of view: technology, people's experience, and society. Anticipation, which involves creativity, and HITLS using appropriate scenarios enable to explore and test possible futures.

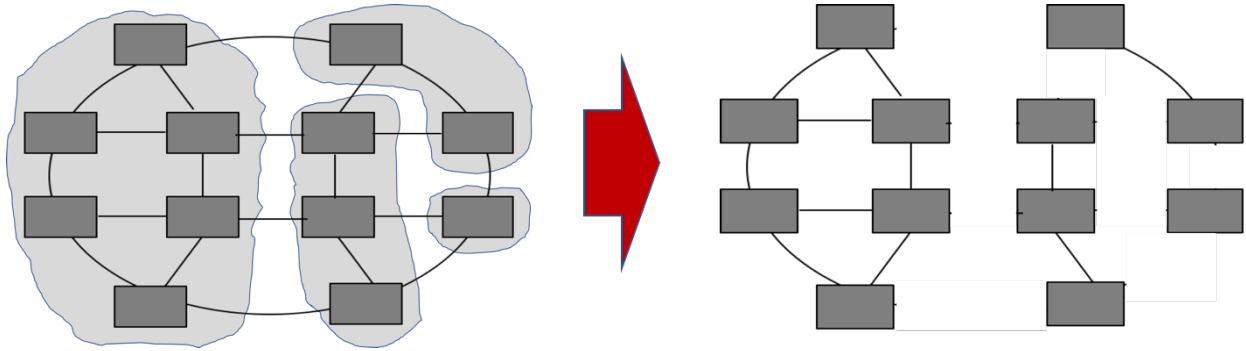

Figure 3. Example of separability property of a system of systems (SoS). Four sub-SoSs of the SoS are separable, and therefore can be analyzed, designed and evaluated in isolation (Boy, 2020).

The lack of consideration for the separability issue in air traffic management is a good example, where, for a long time, most air and ground technologies have been designed and developed in isolation. Recent programs, such as SESAR (Single European Sky ATM Research), try to associate air and ground stakeholders. The TOP model is a good framework for design and development teams to understand and rationalize interdependencies of technology, organizations and people. This objective requires that various activities be observed and analyzed using modeling and simulation, in order to discover emerging properties and functions of the airspace technologies under development, and not await the discovery of these emerging properties and functions at operations time.

## 3. Evolution with respect to models, human roles and disciplines

### 3.1. From single-agent interaction to multi-agent integration

For a long time in aeronautics, we were essentially centered on single-agent interaction with cockpit instruments and controls in the cockpit, and with traffic displays and radio on the ground. Focusing on cockpits, for several decades of the $20^{th}$ century, electrical engineering, computer science and information technology incrementally penetrated commercial aircraft. Many kinds of embedded systems were developed during four steps that Captain Etienne Tarnowski called "the four loops of automation" (Figure 4).

Everything started with automation around the center of gravity, using yoke or side stick and thrust levers. The first loop consisted in a single agent, the autopilot, regulating parameters, such as speed and heading, one parameter at a time. Time constant of the feedback is around 500 milliseconds. Pilots had to adapt to this control loop by changing from the control of flight parameters to supervising the behavior of flight control with respect to a set point.

The guidance loop was developed circa the early eighties. This second feedback loop involves several parameters. Its time constant is around 15 seconds. Note that this feedback loop has been typically implemented on top of the flight control loop. High-level modes of automation appeared and were managed on the flight control unit panel. At the same time, integrated and digital autopilot and auto-throttle were installed.

  5

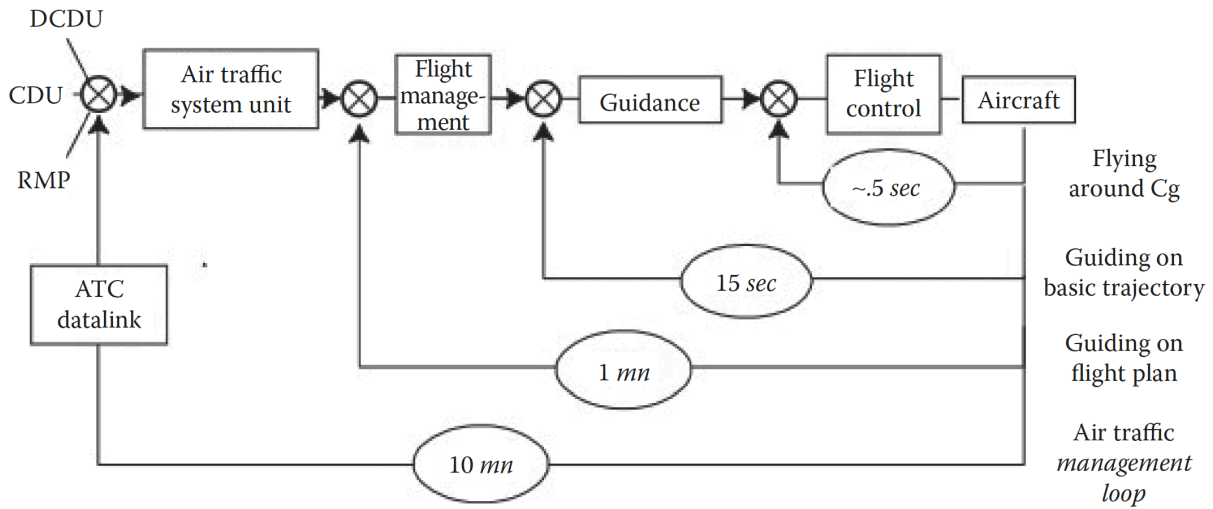

*Figure 4. The four loops of automation of the airspace (Tarnowski, 2006).*

The third loop concerns navigation automation with a time constant of about one minute. Guidance and flight management was integrated circa mid-1980s. This was the first real revolution in the development of avionics systems. We were shifting from control of flight parameters to management of avionics systems, which grew exponentially as software became dominant. For that matter, pilots now have to deal with a variety of avionics systems that qualify as software-based agents, which generate problems as they become more interconnected.

Technology continued to improve and aircraft cockpits became more computerized. As already described elsewhere (Boy, 1998, 2011), commercial aircrafts were equipped with autopilots for a long time (since the 1930s), but what drastically changed is the amount of software in avionics systems. The notion of systems quickly became persistent and pilots' work radically changed from handling flight qualities (manual control) to aircraft systems management. A pilot's role shifted from control to management, exactly like when someone becomes a manager in an organization and has a team of agents to manage. In this case, pilots had to learn how to manage very advanced systems and coordinate their activities. It was not obvious when suddenly a pilot had to become a manager (of systems). This new emerging cognitive function (i.e., systems management), had to be learned and stabilized.

The forth loop concerns air traffic management with a time constant of about ten minutes. We now deal with airspace "automation." This is about air-ground integration. This is a new revolution, where new considerations, such as authority sharing, air traffic complexity management (Hilburn, 2004), and organizational automation (Boy & Grote, 2009). Note that this ATM loop is multi-dimensional and multi-agent.

### 3.2. Systems management and authority sharing

There was a big controversy during the late 1980s when the first highly automated glass cockpits of commercial aircraft were delivered and used (i.e., integration of the first three loops described above). This controversy started with social issues in the beginning of the 1980s because the commercial aircraft industry went from three-crewmen cockpits to two-crewmen cockpits, and downsizing the technical crew was not universally accepted. The role that was previously

 

performed by a flight engineer was shared among the captain, first officer and avionics systems. Function allocation was at stake and we had to find out how to certify these new cockpits. Therefore, we developed human factors methods that enabled the evaluation of aircrew workload and performance, comparing various types of configurations. We needed to demonstrate that a forward-facing cockpit with two crewmembers was as safe as the previous types of cockpits.

At the same time, the number of aircrafts grew exponentially, and induced new issues related to traffic density and airspace capacity. New systems came onboard, such as the Traffic-alert and Collision Avoidance System (TCAS), which provided a new kind of information to the pilot. Not only the TCAS provided traffic alerts, but also advice to climb or descend to avoid a converging aircraft. Not only were the collision avoidance orders from air traffic controllers replaced by the TCAS, but also these orders were, and remain, onboard and given by a machine. Authority shifted from the ground to the aircraft, and from humans to machines. We can see here that technology has resulted in the emergence of a different practice related to a different authority allocation.

The PAUSA[1] project investigated practical issues in authority sharing in the airspace (Boy & Grote, 2009; Boy et al., 2008). In this project, we extensively analyzed the 2002 Überlingen mid-air collision due to a wrong TCAS usage. TCAS introduced a gradual shift from central control to decentralized self-organization (Weyer, 2006). The PAUSA approach to authority was grounded in a multi-agent approach that enables the expression of function distribution, the notion of a common frame of reference, task delegation, and information flows among agents. The metaphor of the shift from the army to an orchestra (Boy, 2009), which represents the ongoing evolution of the airspace multi-agent system, emerged from multiple experience-based investigations within PAUSA.

### 3.3. Human-centered disciplines involved

At this point, it is important to clarify the relationship between the task/activity distinction and distinctions between socio-technical disciplines. More specifically, we need to define the following concepts: task, activity, human factors, ergonomics, Human-Computer Interaction (HCI), HCD and HSI. A task is what is prescribed to human operators or users. An activity is what is effectively performed by human operators or users.

For the last sixty years, socio-technical evolution can be decomposed into three eras in which three communities[2] emerged (Figure 5):

- **Human Factors and Ergonomics** (HFE) was developed after the second world war to correct engineering production, and generated the concepts of human-machine interfaces or user interfaces, and operational procedures; activity-based evaluation could not be holistically performed before products were finished or almost finished, which enormously handicapped possibilities of re-design. Sometimes, activity analyses were carried out prior to designing a new product, based on existing technology and practice; however, this HFE approach forced continuity, reduced risk-taking, and most of the time prevented disruptive innovation.

---

[1] PAUSA is a French acronym for "authority distribution in the aeronautical system". This national project was sponsored by the French Aviation Administration (DPAC) and the aeronautical industry. Nine organizations participated in this project. This work was carried out when the author was the Director of the PAUSA project, at the European Institute of Cognitive Sciences and Engineering (EURISCO).

[2] The author is qualified to talk about these three communities. He is still the Chair of the Aerospace Technical Committee of the International Ergonomics Association (IEA), which encapsulates most HFE societies around the world. From 1995 to 1999, he was the Executive Vice-Chair of the Association for Computing Machinery (ACM) Special Interest Group on Computer Human Interaction (SIGCHI), and Senior Member of the ACM. He is currently Co-Chair of the Human Systems Integration Working Group of the International Council on Systems Engineering (INCOSE).

- **Human-Computer Interaction** (HCI) was developed during the 1980s to better understand and master human interaction with computers; it contributed to the shift from corrective ergonomics to interaction design mainly based on task analysis. Activity-based analysis was introduced within the HCI community by people who understood phenomenology (Winograd & Flores, 1986) and activity theory (Kaptelinin & Nardi, 2006).

- **Human-Systems Integration** (HSI) emerged from the need to officially consider human possibilities and necessities as variables in systems engineering (SE); incrementally combined, SE and HCD lead to HSI, to take care of systems during their whole life cycle (Boy & Narkevicius, 2013). HCD involves looking beyond human factors evaluations and task analyses. It involves activity analysis at design time using virtual prototyping and human-in-the-loop simulations (e.g., we can model and simulate an entire aircraft, fly it as a computing game, and observe pilot's activity). HCD also involves creativity, system thinking, risk taking, prototype development using agile approaches, complexity analyses, organizational design and management, as well as HSI knowledge and skills.

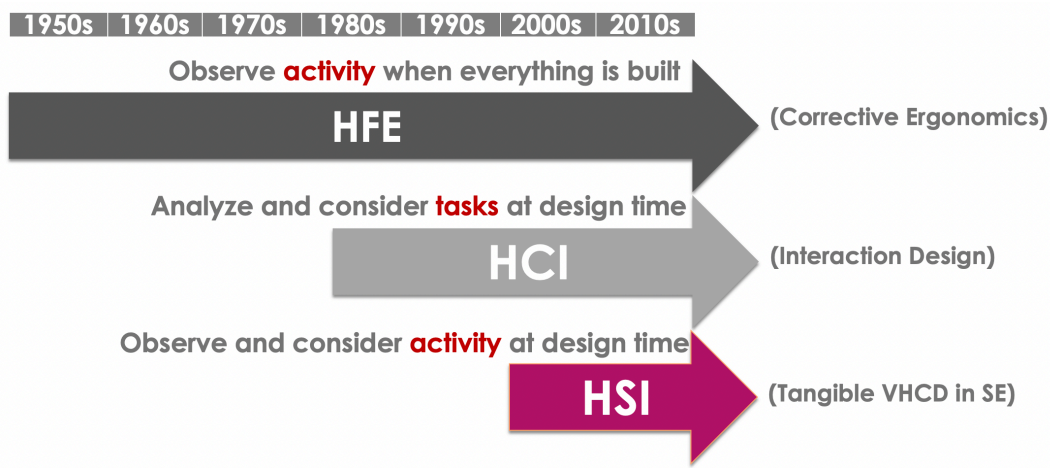

*Figure 5. Human-centered design evolution.*

### 3.4. From automation issues to tangibility issues

During the 1980s and 1990s, automation drawbacks emerged from several HFE studies, such as "ironies of automation" (Bainbridge, 1983), "clumsy automation" (Wiener, 1989[32]), and "automation surprises" (Sarter et al., 1997). These studies considered neither technology maturity nor maturity of practice. Automation can be modeled as cognitive function transfer from people to systems (Boy, 1998). If automation considerably reduced people's burdens, it also caused problems such as complacency, which is an emerging cognitive function (i.e., not predictable at design time, but at operations time).

   8

Today, we can develop an entire aircraft on computers from inception of design to finished product. Therefore, we are able to test its operability from the very beginning, and along its life cycle using HITLS and an agile approach. Consequently, we are able to observe, and therefore analyze, human activity using systems in a simulated environment at design time. The operability of complex systems can then be tested during design. This is the reason why HCD has become a discipline in its own right. HCD enables us to better understand human-systems integration during the design process and then have an impact on requirements before complex systems are concretely developed.

HITLS enables activity analysis at design time, albeit in virtual environments. Even if these environments are very close to the real world, their tangibility must be questioned, and most importantly validated.

What is tangibility? It has two meanings. First, physical tangibility is the property of an object that is physically graspable (i.e., you can touch it, hold it, sense it and so on). Second, figurative tangibility is the property of a concept that is cognitively graspable (i.e., you can understand it, appropriate it, feel it and so on). If I try to win an argument with you, you may argue back, "what you are telling me is not tangible!" This means that you don't believe me; you cannot grasp the concept I am trying to provide. We also may say that you do not have the right mental model to understand it, or that I do not have enough empathy to deliver the message appropriately.

Tangibility is about situational awareness both physically and cognitively. For example, Tan developed the first versions of the Onboard Context-Sensitive Information System (OCSIS) for airline pilots on a tablet PC (Tan, 2015). Physical tangibility considerations led to a better understanding of whether OCSIS should be hand-held or fixed in the cockpit. Other considerations led to the choice of figurative displays of weather visualization going from vertical cylinders to more realistic cloud representations (figurative tangibility). A set of pilots gave their opinions on various kinds of OCSIS tablet configurations. It is interesting to note that the pilots always naturally used the term "tangible" to express their opinions.

Therefore, tangibility metrics should be developed to improve the assessment of complex systems operability. This is where subject matter experts and experienced people enter into play. We absolutely need such people in HCD to help assess HSI tangibility. For example, very realistic commercial aircraft cockpits, professional pilots and realistic scenarios are mandatory to assess tangibility incrementally. OCSIS was tested from the early stages of the design process using human-in-the-loop simulations, by recording what pilots were doing while using it and analyzing their activity. Such formative evaluations lead to system modifications and improvements. HCD is iterative, and agile[3] in the systems engineering sense.

While the 21st century shift from software to hardware is not necessarily straightforward, it is the next dilemma we must address, especially now that we can 3D print virtual systems and transform them into physical systems. We will denote resulting systems, Tangible Interactive Systems (TISs) (Boy, 2016). The TIS concept is very close to the Cyber-Physical Systems (CPSs) concept, which are usually defined as a set of collaborative elements controlling physical entities (Lee, 2008). CPSs are often qualified as Embedded Systems (ESs), but there is a distinction between ESs as purely computational elements, and CPSs as computational and physical elements intimately linked (Wolf, 2014). We can say that CPSs are extended avionics systems. Both TISs and CPSs are strongly based on the multi-agent concept, unlike 20th century automation that was based on the single agent concept. TISs cannot be considered without an organizational approach. More generally, the co-evolution of people's activities and technology necessarily lead to a tangible organizational evolution.

---

[3] The Manifesto for Agile Software Development (http://www.agilemanifesto.org) has been written to improve the development of software. It values individuals and interactions over processes and tools, working software over comprehensive documentation, customer collaboration over contract negotiation, and responding to change (flexibility) over following a plan (rigidity).

## 4. From rigid automation to flexible autonomy

When we know the domain well, both technically and operationally, it is possible to define appropriate procedures that can either be implemented as computer programs (i.e., automation of machines) or followed by human operators (i.e., automation of people). Figure 6 shows how, in a very well-known and/or expected validity space, procedures and automation lead to automation of respectively human and machine functions. Everything goes fine within the validity space (i.e., in nominal situations). However, outside the validity space, the rigidity of both procedures and automation rapidly leads to instability.

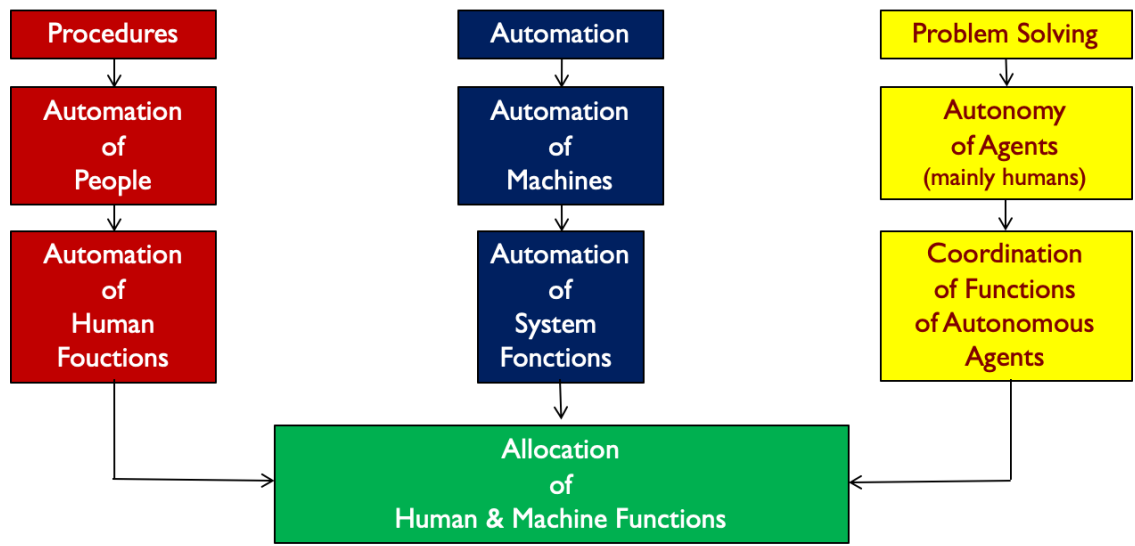

*Figure 6. Procedures, automation and problem-solving leading to the allocation of human and machine functions (Boy, 2020).*

In off-nominal situations (i.e., unexpected, unknown, abnormal or emergency situations), people need to solve problems (Pinet, 2015). Problem solving is a matter of knowledge and knowhow. The more they have such knowledge and knowhow, the more people are autonomous. They also need to have appropriate technological and/or organizational support. Altogether, autonomy is a matter of appropriate technological support enabling flexibility, coordinated organizational support, and people's knowledge and knowhow. Off-nominal situations management involves functions of autonomous human and machine agents that need to be coordinated.

In all cases, functions from automated or autonomous agents need to be correctly allocated to the right agents whether they are people or machines. Such allocation cannot be done entirely statically, but dynamically with the evolution of context.

A fully autonomous agent, whether human or machine, does not exist, or may exist in very limited contexts. Indeed, no agent can always be aware of the overall situation, nor is always capable of making the right decision at the right time, nor act appropriately in the right context. The reason is that there are too many parameters entering into play. Flexible autonomy of an agent will have to be investigated using the following claims:

- An agent is defined as a society of agents (Minsky, 1985);
- Each agent owns appropriate knowledge and processing capabilities, in terms of function(s) and structure(s), to perform a given task;
- The more agents become autonomous in a society of agents, the more coordination rules are needed to keep the stability and sustainability of the overall system.

In other words, agents become increasingly autonomous once they have acquired appropriate knowledge and improved capacities of coordinating with others. The concept of "appropriate knowledge" can be defined as knowledge suitable to have in a given context. The concept of "coordination" can be defined as both prescribed and effective coordination in a given context. Prescribed coordination is a matter of coordinated tasks (i.e., like symphony scores produced by the composer to coordinate the various musical instruments). Effective coordination is a matter of coordinated activity (i.e., a conductor coordinating the various musicians of the orchestra).

Aerospace experience shows the need for deeper support in terms of systemic framework. Over the years, we accumulated embedded systems, now extended to cyber-physical systems, Internet of Things, and other systems. Even if each of these systems can be useful and usable, when they are put together and operated by people, they can become extremely difficult to manage. In critical situations, operational complexity, situation awareness, and workload are directly impacted. Consequently, there is a need for improving HSI. This need leads to a more fundamental requirement, that is defining what a system is really about).

## 5. How software took the lead on hardware

During most parts of the 20$^{th}$ century, hardware and mechanical machines were primary engineering concerns. We built washing machines, cars, aircraft and industrial plants using mechanical engineering methods and tools. At the end of the 20$^{th}$ century, we started to introduce electronics and software into these machines, to the point that we manage to shift practices from mechanical manipulation to human-computer interaction. Computers invaded our lives as mediators between people and mechanical systems. The computer-based User Interface (UI) became a primary issue and solution everywhere in HCI. In aeronautics, we ended up with the concept of interactive cockpits, not because pilots interacted with mechanical things, but because they now interact using pointing devices on computer displays. HFE combined with HCI and cognitive science, as a new discipline often called cognitive engineering, supported machine developments by studying, designing and evaluating UIs, added after a mechanical machine was developed. Software contributed to adding automated functions to machines, with these functions offering new capabilities to people, who had to learn how to interact with automation safely, efficiently and comfortably. In addition, automation came in the form of incrementally-added layers of software that increased situation awareness issues (Endsley, 1996).

It is interesting to notice that, since the beginning of the 21$^{st}$ century, most projects start on computers, with a PowerPoint slide deck, a computing model and visualization, a simulation, and in some cases a computer-based HITLS. For example, the Falcon 7X, developed by Dassault, was entirely built as a giant interconnected piece of software that led to a sophisticated computer game flown by test pilots. For the time, we were able to observe end user activity at design time, even before any piece of hardware was developed. Once such software models and simulation are tested, they enable design teams to develop appropriate requirements for making hardware parts and, in many cases now, 3D printed software models. We have moved from software to hardware, inverting the 20$^{th}$ century's approach that moved from hardware to software. As a result, automation is less of an issue as it was before because HITLS enables testing functional safety, efficiency and usability since the beginning of the design process. The problem is then to better understand the tangibility issue induced by the shift from software models and simulations to the concrete structural world. Tangibility is not only a matter of physics, it is also a matter of intersubjectivity (i.e., mutual understanding) between end users and designers. In other words, end users should be able to understand what machines are doing at appropriate levels of granularity. This is the reason why complexity analysis has become tremendously important in our increasingly interconnected world.

System knowledge, design flexibility and resource commitments are three parameters that should be followed carefully during the whole life cycle of a system. Human systems integration aims to increase the following sufficiently early (Boy, 2020):

- system knowledge, that is knowing about systems at design, development, operations and closeout times, how the overall system, including people and machines, works and behaves;
- design flexibility, that is keeping enough flexibility for systems changes later in development and usages; and
- resource commitments, that is keeping enough "money" for choosing adapted resource management during the whole life cycle of the overall system.

When a technology-centered approach is used (typically what we have done up to now), system knowledge increases slowly in the beginning, growing faster toward the end of the cycle. Design flexibility drops very rapidly, leaving very few alternatives for changes, because resource commitments were too drastic too early during design and development processes.

Instead of developing technology first, software models are used for the development and use of HITLS, which enables activity observation and analysis, and therefore the discovery of emergent properties and functions, that in turn can be considered incrementally in design. This is a typical HCD process. Consequently, system knowledge increases more rapidly during such an agile design and development process. At the same time, design flexibility drops much more slowly, with an inverted concavity, enabling possible changes later in the life cycle. Software-based modeling and HITLS enable testing various kinds of configurations and scenarios, enabling softer resource commitments in the beginning and leaving more comfortable space for appropriate changes.

At this point, we see the need for modeling and simulation for HCD and consequently HSI. If 20th century automation, that is putting software into hardware or transferring human functional knowledge into a machine, brought human functional issues, 21st century tangibilization, that is transforming software models into real-world tangible (concrete) human machine systems is currently raising architectural issues. Before we provide a topological approach to HSI in section 4 of the chapter, let's present and discuss the shift from rigid automation to flexible autonomy.

## 6. Toward a human-centered systemic framework

### 6.1. Systems of systems, physical & cognitive, structures & functions

Historically, engineers used to think about a system as an isolated system, or a quasi-isolated system, which has an input and produces an output (Figure 7). Comparatively an agent, in the artificial intelligence (AI) sense, has sensors and actuators, and a system, in the systems engineering sense, has sensors to acquire an input and actuators to produce an output.

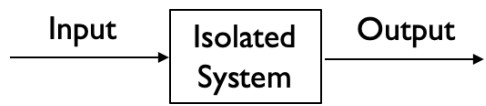

*Figure 7. An isolated system.*

Today, systems are highly interconnected and we talk about a system as an SoS, which means that a system can be represented as an organization of other systems. More generally, a system belongs to a bigger system and is interconnected with other systems (Figure 8). The same holds for agents in AI, which can be organized within a society of agents (Minsky, 1985).

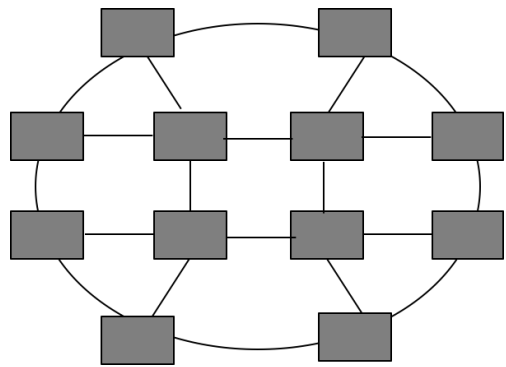

*Figure 8. A system of systems represented as an infrastructure (i.e., a society of agents, or a structure of structures).*

Following up on the separability concept for a system of systems, it is now crucial to have a clear definition of what the system means. A system is a representation of:

1. a human or more generally a natural entity (e.g., a bird, a plant);
2. an organization or a social abstraction (e.g., a team, a community, a law, a legally-defined country, a method); or
3. a machine or a technological entity (e.g., a car, a motorway, a washing machine, a chair).

Consequently, this system definition breaks the traditional meaning of system, conceived as a machine only, but instead encapsulating humans, organizations and machines[4]. Figure 9 presents a simple ontological definition of the "system" conceptual representation.

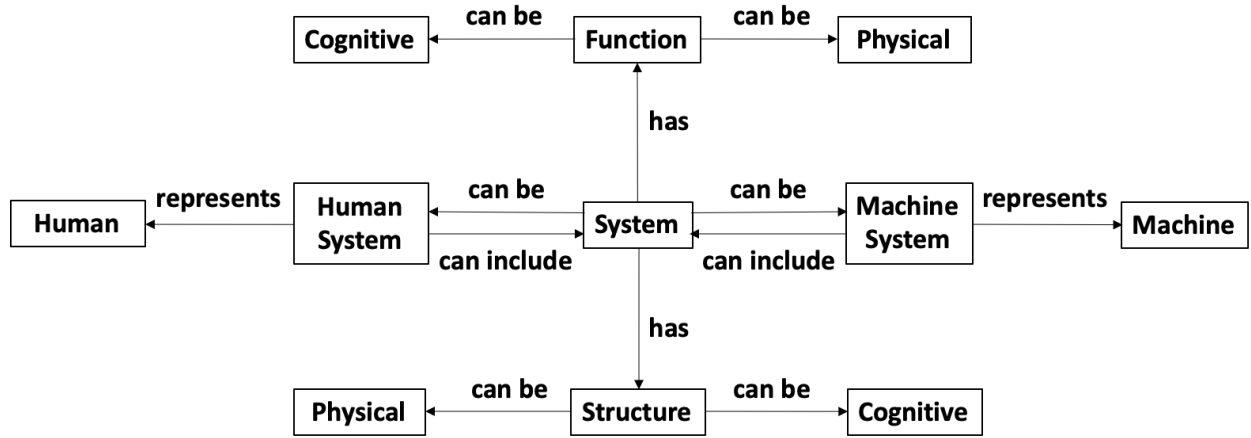

*Figure 9. Synthetic view of the system representation.*

A system can be either cognitive (or conceptual), physical or both (Boy, 2017). It also has at least one structure and one function. Today, machines have software-supported cognitive functions (e.g., the cruise control function on a car enables the car to maintain a set speed). In practice, a system has several structures and several functions articulated within structures of structures and functions of functions. It is interesting to recall the analog definition of an agent in AI provided by Russell and Norvig (2010), which is an architecture (i.e., structure) and a program (i.e., function).

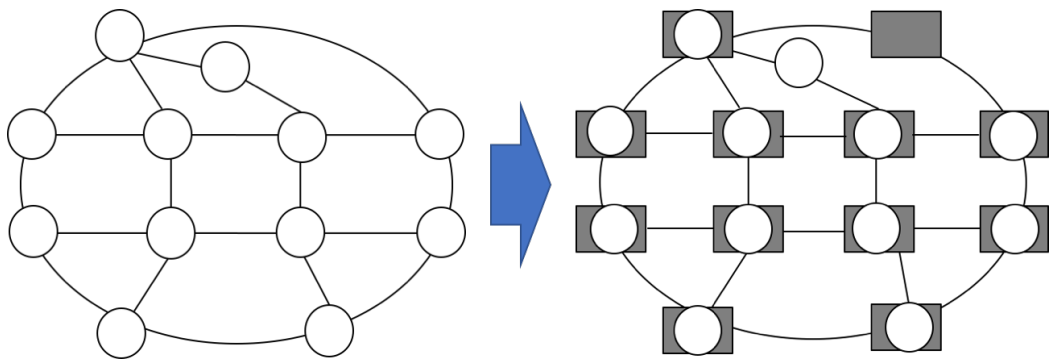

*Figure 10. A function of functions mapped onto a structure of structures.*

Each system is interconnected to other systems either structurally (in terms of a systemic infrastructure) and functionally (in terms of functions appropriately allocated to systems). Summarizing, a system, as a system of systems, is represented by an infrastructure where a network of functions could be dynamically allocated (i.e., a function of functions mapped onto the structure of structures – Figure 10).

---

[4] This systemic view takes Herbert Simon's view of the Science of the Artificial (Simon, 1996), in the sense that he rejected treating human sciences using the exclusive model of the natural sciences (i.e., submission to natural laws) and to break between science and humanities by looking for a common ground that links them. The science of the artificial seeks new constructs that would explain things, which were not previously understood. These artificial constructs could be a language, an ontology, a conceptual model or any kind of representation that makes sense.

### 6.2. Emergent behaviors and properties

At this point, a distinction should be made between deliberately established functions allocated onto an infrastructure and functions that necessarily emerge from system activity. Indeed, systems within a bigger system (i.e., a system of systems) interact with each other to generate an activity. Bertalanffy (1968)[5] said "a system is a set of elements in interaction." Emerging functions are discovered from such activity (Figure 11). The integration of such emergent functions into the system of systems may lead to the generation of additional structures, which we also call emerging structures.

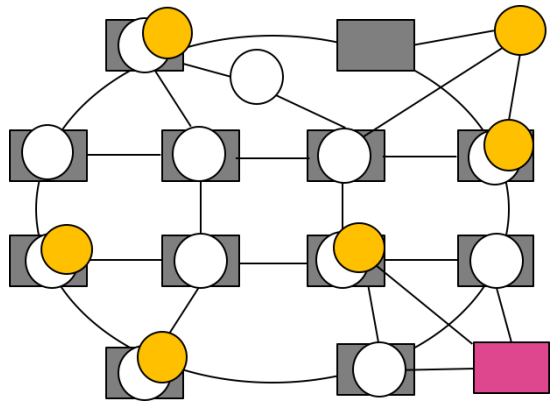

*Figure 11. Emerging functions (yellow) and structures (pink) within an active system of systems.*

### 6.3. Systems of systems properties

The system's purpose is logically defined by its task space (i.e., all tasks the system can perform successfully). Each task is performed by the system using a specific function that produces an activity that can be fully or partially observable (Figure 12).

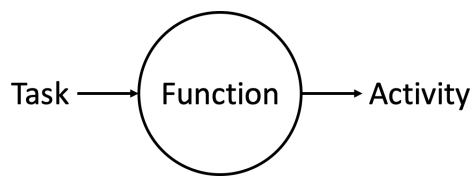

*Figure 12. A function logically transforms a task into an activity.*

A system's function is teleologically defined by three entities:

1. its role within the related system;
2. its context of validity that frames the boundaries of the system's performance; and

---

[5] https://www.sebokwiki.org/wiki/What_is_a_System%3F

3. its set of resources required to perform its role within its context of validity. Resources are systems themselves that have their own cognitive and/or physical functions.

Therefore, according to these definitions, a system can be represented by the recursive schema presented in Figure 13.

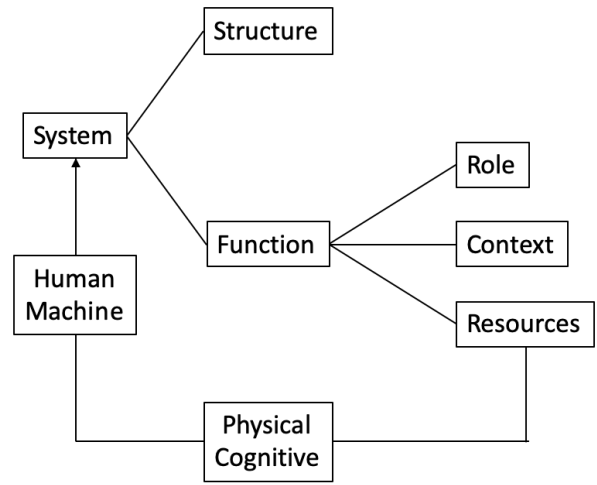

*Figure 13. HSI recursive definition of a system.*

Let us consider a postman represented as a system (or an agent in the AI sense) with the function of delivering letters. The postman as a system is part of a system of systems, which is the postal services. The role of this system is "delivering letters." The context of validity is, for example, seven hours a day five days a week (i.e., a time-wise context in France, for example), and a given neighborhood (i.e., space-wise context). Resources can be physical (e.g., a bicycle and a big bag) and cognitive (e.g., a pattern-matching algorithm that enables the postman to match the name of the street, the number on the door, and the name of the recipient). The corresponding pattern matching algorithm is a cognitive function. Let's consider now that there is a strike, and most postmen are no longer available for delivering letters. Remaining postmen should have longer hours of work in more significant neighborhood until this expansion is so extreme that the postman need helpers to achieve the delivery task successfully.

In this case, a tenure postman should have cognitive resources such as "training", "supervising" and "assessing" temporary personnel. We see that the cognitive function of "delivering letters" owned by a postman (i.e., an agent or a system) has to be decomposed into several other functions allocated to temporary postmen. We start to see an organization developed as an answer to a strike. More generally, a function of functions can be distributed among a structure of structures.

## Conclusion and perspectives

This chapter benefited from years of aerospace experience that contributed to the genesis of a Human Systems Integration (HSI) conceptual framework, useful for the human-centered design of complex systems; where safety, efficiency, and comfort are most required. The HSI experience, concepts, and methods that have been developed in aerospace can be extended to other industrial and public sectors, such as mobility and medicine. More specifically, the concept of the system described in this chapter can be used for the rationalization of HSI in a large variety of domains.

In addition, contemporary technology and organizations are becoming digital, and modeling and simulation will naturally develop over the next few years, providing tremendously useful capabilities to HSI endeavors in engineering design and other processes of the life cycle of systems (i.e., systems engineering).

A question worth asking is: should we continue to talk about human-centered or life-centered technology and organizations? In this chapter, we introduced the first contribution to an HSI ontology (see Figure 10). However, it would be great to expand this approach to life in general, including natural entities and environmental issues. Indeed, aeronautics is increasing in sensitivity to climate change, for example, by trying to find solutions to this major planetary issue. From a general standpoint, we need to address appropriate life-critical constraints and goals for technology design and development (e.g., safety, efficiency, and comfort). More specifically, sustainability should be part of these constraints and goals, including social, economic, and environmental factors.

A shift from the old army pyramidal model to the orchestra model is currently emerging (see the Orchestra model in Boy, 2013). For example, technology and emergent practices have led people to change ways of communicating with each other. The army model induced mostly descendent vertical communication. Transversal communication (e.g., using telephone, email and the Web) contributed to the emergence of the Orchestra model (Boy, 2009; Boy & Grote, 2009). This functional evolution is now changing organizations themselves (i.e., structures). For example, smart phones and the Internet have contributed to change in both industrial and everyday life organizations.

The Orchestra model provides a usable framework for human-systems integration. It requires a definition of a common frame of reference (music theory), as well as jobs such as the ones of human-centered designers and systems architects (composers) who provide coordinated requirements (scores), highly competent socio-technical managers (conductors) and performers (musicians), and well-identified end-users and engaged stakeholders (audience). Having this organizational model in mind, it is now crucial to use it in HCD. More specifically, in the framework of this chapter, it can be very useful for ATM research and design.

It is time to further develop methods and tools for the integration of people and organizations in the life-critical design and development of new technology. To do this, HSI foundations need to be further developed to support human-centered design. This chapter presented conceptual solutions to this endeavor. More is to come, and HSI research needs to be promoted and supported.

# References

Bainbridge, L. (1983). Ironies of Automation. *Automatica*, International Federation of Automatic Control, Pergamon Press, Vol. 19, No. 6, pp. 775-779.

Billings, C.E. (1997). *Aviation Automation. The Search for a Human-Centered Approach*. Lawrence Erlbaum Associates: Mahwah, NJ.

Boy, G.A. (1993). Human-Computer Interaction in the Cockpit. Invited lecture at Inter-CHI'93, Amsterdam, NL.© Guy A. Boy (2020). Aerospace Human System Integration Evolution over the Last 40 Years. In A Framework for Human System Engineering Applications and Case Studies, H.A.H. Handley & A., Tolk (Eds.). IEEE Press, Wiley, USA. ISBN-13: 978-1119698753. Preprint.

Index